\documentclass[apjl]{emulateapj}

\usepackage{graphicx}
\shortauthors{Martig \& Bournaud}
\shorttitle{Formation of late-type spiral galaxies}

\begin{document}

\title{Formation of late-type spiral galaxies:\\
gas return from stellar populations regulates disk destruction and bulge growth}

\author{Marie Martig and Fr\'ed\'eric Bournaud}
\affil{CEA-Saclay, IRFU, SAp, 91191 Gif-sur-Yvette, France.}
\email{marie.martig@cea.fr}

\begin{abstract}
Spiral galaxies have most of their stellar mass in a large rotating disk, and only a modest fraction in a central spheroidal bulge. This challenges present models of galaxy formation: galaxies form at the centre of dark matter halos through a combination of hierarchical merging and gas accretion along cold streams.  Cosmological simulations thus predict galaxies to rapidly grow their bulge through mergers and instabilities, and to end-up with most of their mass in the bulge and an angular momentum much below the observed level, except in dwarf galaxies. We propose that the continuous return of gas by stellar populations over cosmic times could help to solve this issue. A population of stars formed at a given instant typically returns half of its initial mass in the form of gas over 10 billion years, and the process is not dominated by supernovae explosions but by the long-term mass-loss from low- and intermediate-mass stars. Using simulations of galaxy formation, we show that this gas recycling can strongly affect the structural evolution of massive galaxies, potentially solving the bulge fraction issue, as the bulge-to-disk ratio of a massive galaxy can be divided by a factor of 3. The continuous recycling of baryons through star formation and stellar mass loss helps the growth of disks and their survival to interactions and mergers. Instead of forming only early-type, spheroid-dominated galaxies (S0 and ellipticals), the standard cosmological model can successfully account for massive late-type, disk-dominated spiral galaxies (Sb-Sc).
\end{abstract}


\section{Introduction} 
The formation of disk-dominated galaxies remains a challenge for modern cosmology \citep{white09, burkert09}. Bright spiral galaxies with stellar masses around $10^{11}$~M$_{\sun}$ at $z=0$ typically have bulge-to-disk mass ratios (B/D) of 0.2 to 0.4 \citep{weinzirl}. For instance, the Milky Way itself has a B/D ratio of 0.35-0.40 \citep{robin03}. Luminosity ratios in near-infrared and optical bands are even lower \citep{Graham2008}.  However, cosmological simulations predict much higher bulge fractions, and consequently a baryonic angular momentum much lower than observed. The hierarchical assembly of dark matter halos drives successive galaxy mergers, followed by a rapid growth of central bulges. Indeed, a single major merger, even starting with a high gas fraction, will generally end-up with more baryons in the bulge than in the disk \citep{SH05,robertson06,hopkins09b}. Successive minor mergers also drive bulge growth (Bournaud, Jog \& Combes 2007). 

It has been realized recently that the assembly of baryons onto galaxies is far from being only driven by mergers. Rapid accretion of cold gas is another major mode of galaxy assembly \citep{dekel09}. This cold accretion mode could apparently feed disk-dominated galaxies, but it makes the disks so massive and turbulent that they violently fragment into giant clumps (Bournaud \& Elmegreen 2009; Dekel, Sari \& Ceverino 2009; Agertz, Teyssier \& Moore 2009; Burkert et al. 2009). Clump coalescence will rapidly fuel the bulge (\citealp{noguchi}; Bournaud, Elmegreen \& Elmegreen 2007), and signatures of this additional bulge formation process have been observed \citep{genzel08, elmegreen09}. Disk internal secular evolution can also lead to the slow buildup of pseudo-bulges \citep{Kormendy2004}.

A general result is that the rapid formation of massive bulges appears unavoidable because of the combination of mergers and disk instabilities. The most disky galaxies in cosmological models have B/D around 1 in the best cases and generally higher, in both the merger-driven and stream-fed dominant modes (\citealp[e.g.,][]{governato09,white09,gibson09,scannapieco09}; Ceverino, Dekel \& Bournaud 2009). The bulge fraction problem is related to the well-known spin crisis: galaxies could gain angular momentum from gravitational torquing by satellites (Porciani, Dekel \& Hoffman 2002), but the early build-up of bulges quenches this process. Energy feedback from supernovae explosions can regulate star formation and keep the gas fraction higher, so that disk components survive mergers more easily. Stream-fed disks with efficient supernovae feedback can then have realistic rotation velocities, in agreement with the observed Tully-Fisher relation \citep{maller-dekel, scannapieco08,CK09,governato09,gibson09,Piontek2009a}, but the mass fraction in these fastly rotating disks remains too low compared to their bulges, meaning that the real angular momentum is low. High bulge fractions thus remain as a separate issue that supernovae feedback cannot solve, even when many feedback models are tested in numerical simulations \citep{scannapieco09,Piontek2009b}. Apart for dwarf galaxies \citep{governato09b}, the standard models apparently fail to form high-spin, disk-dominated galaxies.

A stellar evolution process that is more general than supernovae feedback is the continuous return of gas from stars of any mass, in particular through stellar winds and planetary nebulae \citep{faber76}. The gas return fraction of a coeval stellar population over the Hubble time can be as high as 50\% for a Chabrier IMF, or around 40\% for a Scalo IMF, with small dependencies on the metallicity (Jungwiert, Combes \& Palou\v{s} 2001; Lia, Portinari \& Carraro 2002; Pozzetti et al. 2007).
Supernovae account only for about 10\% of this return fraction, which is dominated by low- and intermediate-mass stars. Half of the stellar mass forming at redshift $ z \sim 2$ will thus be returned in the form of gas by redshift 0, which can cool down in the disk and form new stars that will in turn return gas, in a continuous recycling process. This is another type of feedback, which is more a {\em mass} feedback than the energy feedback from supernovae explosions usually considered.  This continuous mass return has already been included in a few cosmological simulations \citep[e.g.,][]{SC02,Wiersma2009}, but these simulations were limited to high redshift, or lacked resolution to clearly resolve disks and bulges, or the specific effects of gas return were not explored.

Here we show that this known process has unsuspected effects on regulating the growth of bulges and the survival of massive disks in spiral galaxies. Using a cosmological simulation of the formation of a massive galaxy, which in the standard model ends-up with a B/D ratio larger than unity, we show that continuous gas recycling reduces the final B/D ratio to 40\%, and the stellar light fraction in the bulge by a factor of 3. The mass and disk rotation speed of this galaxy are consistent with observed scaling relations, and its remarkably flat rotation curve is typical for late-type, high-spin spirals. The mass return of stars regulates the bulge growth and angular momentum dissipation in such proportion that a late-type spiral (Sb-Sc) is formed instead of an early-type galaxy. This shows how strongly stellar evolution can affect the formation of galaxies, and help to solve the issue of massive late-type spiral galaxy formation.

\section{Simulation}
\begin{figure}
\centering
\includegraphics[width=8cm]{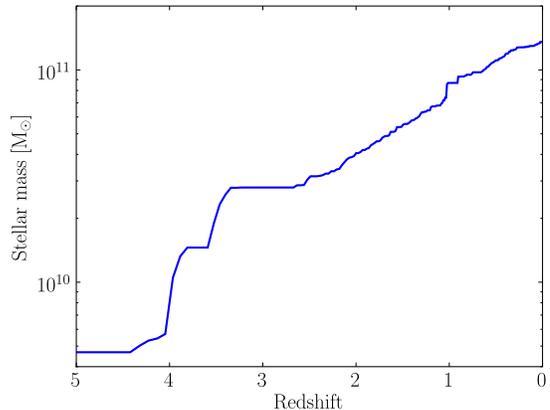}
\caption{Stellar mass evolution of the studied galaxy as a function of redshift. An intense phase of mergers takes place between $z=5$ and $z=3$. The subsequent evolution is much smoother, dominated by diffuse gas infall: the most important merger of this second phase takes place at $z=0.8$ with a mass ratio of 6:1.}\label{evol_mass}
\end{figure}

Our model is based on a high resolution re-simulation of a galaxy identified in a large-scale cosmological simulation. The initial cosmological run was performed assuming standard $\Lambda$-CDM cosmological parameters using the RAMSES code \citep{teyssier02} and is described in \cite{martig09}. The chosen dark matter halo  has a virial mass of 1.3$\times 10^{12}$ M$_{\sun}$ at redshift 0, in a field environment. We avoid dense environments and halos undergoing many violent mergers so that the chosen halo is prone to hosting a disk-dominated spiral galaxy at $z=0$. The mass assembly history of the re-simulated galaxy is shown on Figure~\ref{evol_mass}, and results from a combination of galaxy mergers and diffuse infall from cosmic gas filaments.

Using the zoom-in technique presented in \cite{martig09}, we re-simulate the evolution of the chosen galaxy from $z=5$ to $z=0$ inside a 400~kpc-large zoom area. The spatial resolution (gravitational softening) is 150~pc and the mass resolution (particle mass) is $1.5\times 10^4$~M$_{\sun}$ for gas, $7.5\times 10^4$~M$_{\sun}$ for stars, and  $3\times 10^5$~M$_{\sun}$ for dark matter. Gas dynamics is modeled with a sticky-particle scheme and star formation is computed with a Schmidt-Kennicutt law with an exponent of 1.5. The threshold for star formation is set to 0.03 M$_{\sun}$pc$^{-3}$, which corresponds to the minimal density for diffuse atomic clouds formation \citep{Elmegreen2002}.

\begin{figure*}
\centering
\includegraphics[width=\textwidth]{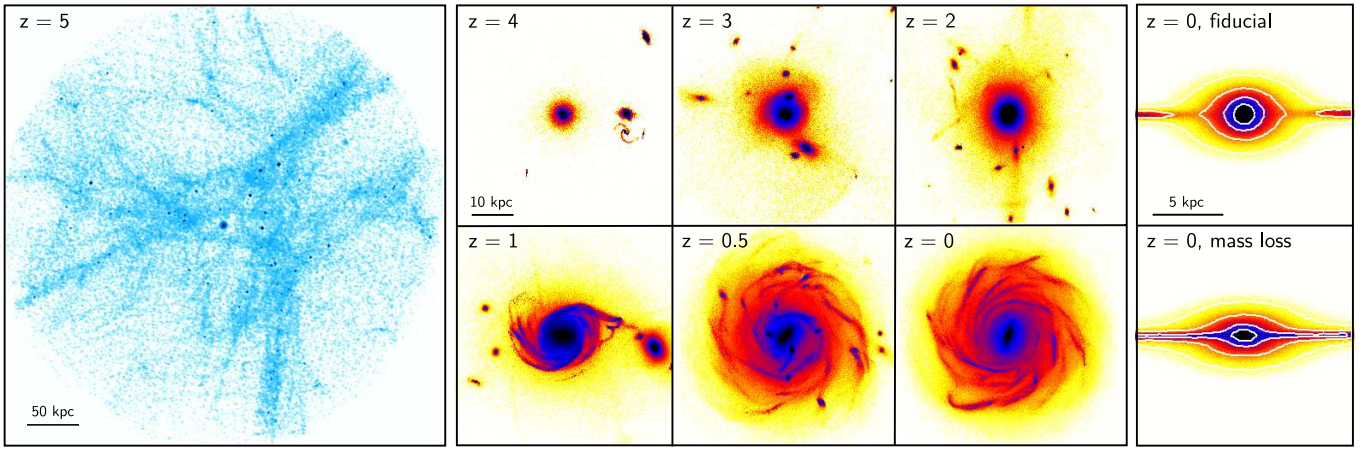}
\caption{Time evolution of the studied galaxy. {\bf Left:} large scale gas distribution at $z=5$. The main galaxy is seen at the center; the box side is 400~kpc. {\bf Middle:} evolution from $z=4$ to $z=0$ of the face-on $i$-band surface brightness, with stellar mass loss (snapshot size 50$\times$50~kpc, log scale). {\bf Right:} Edge-on views of the $i$-band surface brightness at $z=0$ in log scale.
Each snapshot is 15$\times$15~kpc, with contours at 17.25, 18.5, 19.75 and 21~mag~arcsec$^{-2}$. In both simulations, a spheroid is first formed by successive mergers at $z>2$. A large spiral disk is built at lower redshift, both from gas found in-situ and accreted from cosmic filaments and companion galaxies. Stellar mass loss supports disk survival in mergers and disk rebuilding after mergers, leading to a much lower B/D. At redshift 0, a late-type, disk-dominated Sb-Sc galaxy with a modest bulge is formed, instead of an early-type S0a-like system with a large and massive bulge and a low-mass disk.}\vskip 0.5cm \label{snapshots}
\end{figure*}

The galaxy is initialized at $z=5$ as a very gas-rich disk, with a gas fraction of 0.5 with respect to total baryonic mass. The initial dark halo follows a Burkert profile with a core radius of 5.9~kpc and a truncation radius of 13.8~kpc. Gas and stars are in a disk with exponential scale-lengths of 540 and 230~pc, respectively. Starting the simulation at $z=5$ captures 97\% of the mass assembly history of this galaxy, as the total mass (including dark matter) at redshift 5 is only $3.7 \times 10^{10}$ M$_{\sun}$. For this reason and because numerous mergers rapidly destroy the proto-disk into a spheroid around which the regular bulge and disk components will gradually grow (Figure~\ref{snapshots}), our choice of starting with a disk is not influential in the final $z=0$ properties of the galaxy.

We do not include the energy feedback from supernovae. There is no unique model for this, and feedback is anyway most efficient in dwarf galaxies where the structure is strongly influenced \citep{governato09b}; the effects in massive galaxies are more moderate \cite[e.g., ][]{agertz10}. Not including feedback would be a serious problem at very high redshift ($z>5$) because of the overcooling issue \citep{SommerLarsen}, or in very high resolution (pc-scale) simulations to avoid the collapse of all gas into very dense clumps that would never be disrupted. This is not resolved in our simulations at resolutions of 100~pc or more. At these scales, feedback mostly regulates star formation, and we calibrate the star formation efficiency in our simulations so that the gas consumption time matches typical observed values, so this effect of feedback is indirectly modeled.

In turn, we implement the continuous gas return from high-, intermediate- and low-mass stars. To this aim, we follow the scheme proposed by \cite{jungwiert01}: each stellar particle represents a population of stars born at the same time, whose return rate is given by:
\begin{equation}
\dot{M} = - M \frac{C}{t-T}
\end{equation}
where $t$ is the age of the particle since its formation, $T$ is set to 4.86~Myr to fit standard IMF return function \citep{jungwiert01}, and $C$ is such that the integrated return fraction of a stellar population over 10~Gyr is 40\%. The returned gas is directly set cold, immediately available for star formation. Real stellar winds would provide warmer gas, but overestimating the delay before it can form new stars would artificially increase the gas fraction and the capacity of disks to survive mergers: to avoid such favorable biases, we let the returned gas immediately available for star formation.

\medskip

We performed a ``fiducial simulation'' without gas return from evolved stars, and a simulation with stellar mass loss. The structural properties were analyzed $z=0$ in both cases. We measured bulge-to-disk mass ratios with different techniques. First, we used azimuthally-averaged stellar density profiles: an exponential disk is fitted and the mass excess in the central regions corresponds to the bulge. We also performed a decomposition based on kinematics, and more precisely on the angular momentum of each star. We computed the total angular momentum of the gas disk in its inner 10~kpc, and set this as the $z$ axis. For each stellar particle, we then computed $\epsilon=j_z/j_{circ}$, the ratio of its angular momentum along the $z$ axis to the angular momentum it would have if it were on a circular orbit (that is $j_{circ}=r\times v_{circ}$). The distribution of the values of $\epsilon$ shows two peaks: one around $\epsilon=0$, that corresponds to bulge stars, and one around $\epsilon=1$, that corresponds to disk stars. The limit between bulge and disk is set at the value of $\epsilon$ corresponding to the minimum of the distribution between these two peaks.

 Finally, we also computed $i$-band surface luminosities using the spectral evolution model PEGASE.2 \citep{Fioc1999} to derive bulge-to-disk ratios in the $i$-band, assuming a solar metallicity: the $i$-band mass-to-luminosity does not vary much across the radial metallicity gradient of typical $z=0$ spiral galaxies.

\section{Results}

The time evolution of the studied galaxy is shown on Figure~\ref{snapshots}. Interactions and mergers with companion galaxies rapidly transform the initial disk into a spheroid. At later times, star formation in gas left over after these early mergers and gas gradually brought in by cosmic gas flows and merging satellites builds a large disk, while interactions continue to grow a bulge - at a rate that depends on the model. At $z=0$, both cases have a central bulge and an exponential disk with grand design spiral arms. A major difference, though, lies in the distribution of the $z=0$ stellar mass in the disk and bulge components. 

\begin{figure}
\centering
\includegraphics[width=8cm]{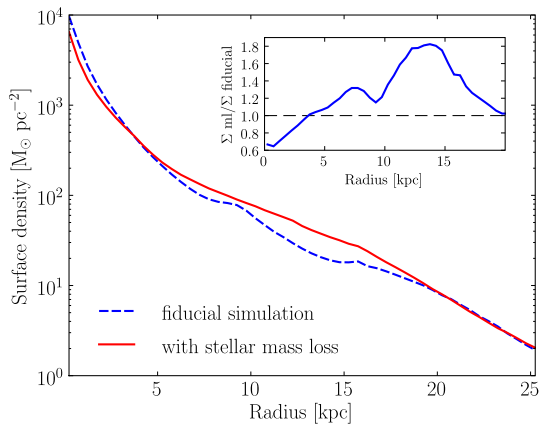}
\caption{Stellar surface density profiles for the fiducial simulation and the one including stellar mass loss. Both galaxies have a central bulge and an exponential disk, but their relative importance differs. On the inset is shown the ratio between the stellar surface density when stellar mass loss is included to the surface density in the fiducial simulation: stellar mass loss reduces the bulge density and increases the disk one, thus decreasing B/D. Differences are seen even on {\it log-scale} radial density profiles.}\label{profiles}
\end{figure}

Without stellar mass-loss, the bulge to disk mass ratio derived from the stellar mass profile is slightly above 1, which is typical for massive ``disk'' galaxies in current $\Lambda$-CDM models (see Introduction). The disk component contains barely half of the stellar mass, even though we include both the thin and thick disks.
 When the continuous stellar mass loss is included, the bulge to disk mass ratio is reduced to 0.40: the bulge is less massive, with a lower density and a smaller size (right panels of Figure~\ref{snapshots}), while the disk is denser with a similar scale length (Figure~\ref{profiles} and Table~\ref{Tab}).

\begin{figure}
\centering
\includegraphics[width=8cm]{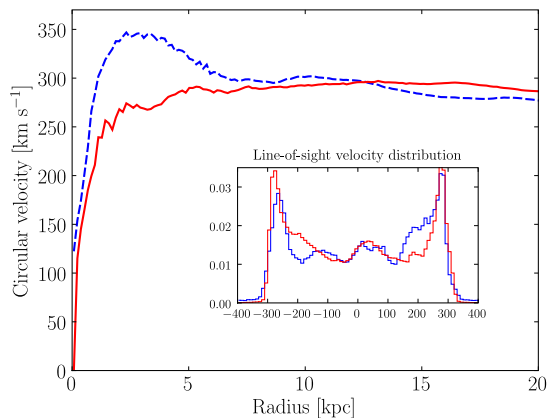}
\includegraphics[width=8cm]{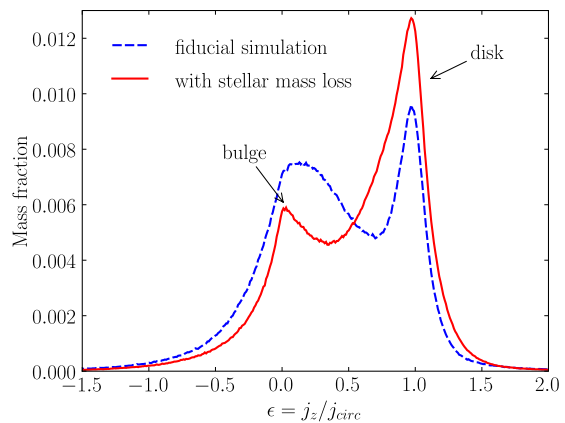}
\caption{Kinematics in the two simulations. The top panel shows the rotation curves. The prominent bulge in the fiducial run causes a central bump in the circular velocity profile, while this profile is much flatter for the simulation including stellar mass loss. On the inset is shown the radial velocity distribution for the galaxy seen edge-on in both simulations: the full width at half maximum is used for comparisons with observations of the Tully-Fisher relation. The bottom panel shows the distribution of the ratio of the angular momentum of the stars along the $z$ axis to the angular momentum they would have on circular orbits: bulge stars are found around $\epsilon=0$ while disk stars are found around $\epsilon=1$. The bulge component is significantly reduced under the effect of stellar mass loss.\label{rotcurve}}
\end{figure}

Other definitions of B/D lead to the same conclusion. A kinematical decomposition even shows a more drastic effect, with B/D decreasing from 1.40 to 0.32 (Figure~\ref{rotcurve} and Table~\ref{Tab}). In the $i$ band luminosity, B/D decreases from 0.49 to 0.16. Similar values were obtained when the simulation was run with a twice lower spatial resolution and a eight times higher particle mass, suggesting reasonable convergence of the simulation.

\begin{table*}
\centering
 \caption{\label{Tab} Disk and bulge properties at $z=0$ with and without stellar mass loss}
\begin{tabular}{ccc}
\hline \noalign{\smallskip} 
 & Fiducial run & With mass loss \\
\hline \noalign{\smallskip} 
Stellar disk mass (M$_{\sun}$)& $6.1\times 10^{10}$& $8.8\times 10^{10}$\\
B/D (stellar mass profile) & 1.01 &0.40\\
B/D (kinematics) & 1.40 &0.32\\
B/D ($i$ band) & 0.49 & 0.16\\
Disk exponential scale length (kpc) & 4.5&  5.0 \\
Optical radius $R_{25}$ (kpc) & 22.0 &  19.8 \\
Bulge half-mass radius (kpc) &  1.3&  1.6\\
Absolute $I$ band magnitude & -22.8 & -22.9 \\
HI linewidth (km s$^{-1}$) & 583 & 591 \\
\hline \noalign{\smallskip} 
\end{tabular}\vspace{.25cm}
\end{table*}

The effect of stellar mass loss on the $z=0$ structural properties of the galaxy is thus major, although we studied quite a massive galaxy for which preserving disks is harder than in lower-mass systems. Continuous stellar mass loss transforms a bulge-dominated system, typically classified as S0/Sa in the Hubble classification \citep{Graham2008} into a disk-dominated Sb/Sc-like galaxy, with a bulge fraction typical for the Milky Way and the majority of massive spirals in the nearby Universe. 

\medskip

As a result from the different mass distribution, the rotation curve at $z=0$ has a different profile (Figure~\ref{rotcurve}). In the fiducial case, the rotation curve is peaked a low radii before decreasing and reaching a flat plateau. This is typical for bulge-dominated, S0-Sa galaxies (e.g., M31 -- Chemin, Carignan \& Foster 2009; NGC 3031, 4736, or 2841 in \citealp{deblok2009}). The regulation of bulge growth by continuous stellar mass-loss results in a flater rotation curve, which is typical of disk-dominated spiral galaxies like NGC~2903, 3198, 3621 \citep{deblok2009}. The comparison of rotation curve with typical cases also corresponds to a shift by two categories in the Hubble classification scheme, from early-type disky galaxies to late-type disk-dominated galaxies, in agreement with the direct estimates of bulge fractions.

We also derived the equivalent of an observed atomic gas (HI) linewidth: we select gas in regions denser than one atom per cubic centimeter in order to include the warm/cold gas disk without the hot halo, simulate the velocity spectrum for the galaxy, and compute its FWHM averaged on several edge-on projections.  The $I$~band absolute magnitude to HI linewidth ratio (see Table \ref{Tab}) puts both systems very close to the observed Tully-Fisher relation for local spiral galaxies as observed by \cite{Springob2007}. The Tully-Fisher relation is nevertheless not the best direct estimate of the baryonic angular momentum of a galaxy since it measures the disk rotation speed at large radii without indicating whether or not a large fraction of the stellar mass lies in a concentrated, low angular momentum bulge.

A similar effect has been also found in two other simulations. We measured bulge and disk mass following the decomposition based on kinematics. In one case (a galaxy with M$_{*}=8\times10^{9}$~M$_{\odot}$ at $z=5$ and 1.3$\times10^{11}$~M$_{\odot}$ at $z=0$), B/D decreases from 1.24 to 0.75 when mass loss is included, in the other case (with M$_{*}=5.3\times10^{9}$~M$_{\odot}$ at $z=5$ and $4\times10^{10}$~M$_{\odot}$ at $z=0$), it decreases from 0.94 to 0.49.
\section{Conclusion}

Our simulation of cosmological galaxy formation shows that the continuous gas return from high-, intermediate- and low-mass stars over cosmic times can strongly regulate the growth of bulge and the dissipation of angular momentum in disk galaxies. There are at least two ways in which stellar mass loss can support the survival of large, massive disks. First, it can keep the gas fraction higher in young galaxies before a merger occurs, which helps disks survival against destruction into bulges \citep{hopkins09a}. Second, the bulge, stellar halo and thickened disk components release fresh gas after interactions and mergers: this decreases their mass and increases that of the thin gas disk. The disk re-formed by gas returned by a bulge or halo should have a low angular momentum, but can be torqued by satellites into a large, high-spin disk that gradually forms new stars (an example of such an interaction is shown at $z=1$ on Figure~\ref{snapshots}). The radial migration of disk stars can also arise from resonant interactions with spiral arms: \cite{Roskar2008} have shown, in a Milky-Way like galaxy, that up to 50\% of stars in the solar neighborhood could have migrated from the inner disk. In our model, the large disk scale length is preserved by the combination of stellar mass-loss and radial migration following external interactions and/or internal evolution.

Overall, the effect of stellar mass loss on disk survival and regulation of bulge growth is major. It reduces the bulge-to-disk ratio by a factor up to $\sim$ 3 in both the stellar mass and light. This factor is what was typically missing in cosmological models to account for the formation of massive, disk-dominated, late-type spiral galaxies.

This does not imply that all galaxies would end-up with a spiral-like morphology at redshift zero, and we briefly presented cases of galaxies with other merger and star formation histories where the decrease of B/D is weaker. Gas return promotes disk survival against mergers and disk instabilities, but will preserve a disk-dominated galaxy only if the mergers are not too numerous, do not happen too late, and the internal instabilities not too violent. For instance, a galaxy undergoing a major merger at low redshift would still end-up as a $z=0$ elliptical or lenticular, because of a stellar population globally too old to return a significant amount of gas. Such $z=0$ ellipticals are indeed found in our cosmological simulations with the same technique \citep[e.g. in][]{martig09}. Gas consumption, stripping and strangulation in groups, will also support the survival of early-type elliptical and S0s. Thus, gas return from stellar populations can explain the origin of late-type galaxies, without challenging the formation of early-type galaxies in systems that undergo later mergers or more violent internal instabilities.

Other sorts of feedback processes, like the energy released by supernovae or the radiation pressure from young massive stars (Murray, Quataert \& Thompson 2009) could further regulate the bulge growth especially in low-mass galaxies, potentially forming almost bulgeless galaxies. However, we propose that the formation of massive late-type galaxies in the $\Lambda$-CDM Universe is explained not by an unknown combination of minor factors, but mostly by one major effect, namely by the gas return from stellar populations all across their initial mass function. This process can largely help to solve one of the main challenges in the origin of modern galaxies, namely the origin of late-type, disk-dominated spiral galaxies like the Milky Way.

\acknowledgements

We acknowledge support from GENCI-CCRT (Grant 2009-042192) and Agence 
Nationale de la Recherche (ANR-08-BLAN-0274-01), and help from Romain Teyssier with the re-simulation technique.

\end{document}